\pdfoutput=1

\documentclass[11pt,twoside,a4paper,cmspaper,final,collab]{cms-tdr}
\def\svnVersion{177388}\def\cmsCernNo{2012-096}\def\cmsCernDate{\today}\def\cmsMessage{Submitted to Physical Review Letters}

\begin{document}\cmsNoteHeader{HIN-11-012}

\hyphenation{had-ron-i-za-tion}
\hyphenation{cal-or-i-me-ter}
\hyphenation{de-vices}

\RCS$Revision: 122833 $
\RCS$HeadURL: svn+ssh://svn.cern.ch/reps/tdr2/papers/HIN-11-012/trunk/HIN-11-012.tex $
\RCS$Id: HIN-11-012.tex 122833 2012-05-19 17:06:48Z ssanders $
\newcommand {\roots}    {\ensuremath{\sqrt{s}}}
\newcommand {\rootsNN}  {\ensuremath{\sqrt{s_{_{NN}}}}}
\newcommand {\npart}    {\ensuremath{N_\text{part}}}
\cmsNoteHeader{HIN-11-012} 
\title{Azimuthal anisotropy of charged particles at high transverse momenta
in PbPb collisions at \texorpdfstring{\rootsNN\ = 2.76\TeV}{sqrt(s[NN]) = 2.76 TeV}}

\date{\today}

\abstract{
The azimuthal anisotropy of charged particles in PbPb collisions at \rootsNN\ = 2.76\TeV
is measured with the CMS detector at the LHC
over an extended transverse momentum (\pt) range up to
approximately 60\GeVc. The data
cover both the low-\pt region associated with hydrodynamic flow phenomena
and the high-\pt region where the anisotropies may
reflect the path-length dependence of parton energy loss in the created medium.
The anisotropy parameter ($v_{2}$) of the particles
is extracted by correlating charged tracks with respect to the
event plane reconstructed using the energy deposited in forward-angle
calorimeters.
For the six bins of collision centrality studied, spanning the range of 0-60\% most-central events,
the observed $v_{2}$ values are found to first increase with \pt, reaching a maximum
around \pt\ = 3\GeVc and then to gradually decrease to almost zero, with the decline
persisting up to at least \pt\ = 40\GeVc over the full centrality
range measured.
}

\hypersetup{%
pdfauthor={CMS Collaboration},%
pdftitle={Azimuthal anisotropy of charged particles at high transverse momenta in PbPb collisions at sqrt(s[NN]) = 2.76 TeV},%
pdfsubject={CMS},%
pdfkeywords={CMS, physics, heavy ions, elliptic flow}}

\maketitle 

The experiments at the Relativistic Heavy Ion Collider (RHIC)
have provided evidence for the formation of a strongly-coupled
quantum chromodynamics (QCD) state of matter in
ultra-relativistic nucleus-nucleus
interactions~\cite{BRAHMS,PHENIX,PHOBOS,STAR}.
The opaqueness of this matter to high-energy partons leads to
a ``jet quenching'' phenomenon where the final-state particle
yield at high transverse momentum (\pt) is found suppressed 
compared to that expected from
the scaling of pp collision yields~\cite{BRAHMS,PHENIX,PHOBOS,STAR,Aamodt:2010jd,cms:2012nt}.
A large energy loss for partons traversing
the dense QCD medium is also suggested by 
the recent observation of a large momentum imbalance of reconstructed
back-to-back jets~\cite{Aad:2010bu,Chatrchyan:2011sx,Chatrchyan:2012ni}
in PbPb collisions at
the Large Hadron Collider (LHC).

Despite the progress made on the theoretical description
of jet quenching in the past decade~\cite{Majumder:2010qh},
some of its key properties, such as the detailed path-length 
dependence of parton energy loss,
remain unknown. In addition to  measurements of hadron-yield suppression, 
observables such as
the azimuthal anisotropy of high-\pt\ hadrons are 
needed to differentiate between the 
theoretical approaches~\cite{Peigne:2008wu,Wicks:2005gt,Jia:2010ee,
Jia:2011pi,Renk:2010mf,Betz:2011tu}.
The anisotropy can be characterized by the second-order 
Fourier harmonic coefficient ($v_{2}$) in the 
azimuthal angle ($\phi$) distribution of the
hadron yield, $dN/d\phi \propto 1+2 v_{2}\cos(2(\phi-\Psi_\mathrm{PP}))$,
where $\Psi_\mathrm{PP}$ is the event-by-event azimuthal angle
of the ``participant plane." In a non-central heavy-ion collision,
the overlap region of the two colliding nuclei has a lenticular shape
and the interacting nucleons in this region are known as ``participants.''
The ``participant plane'' is defined by the beam direction
and the short direction of the lenticular region. In general, the
participant plane will not contain the reaction impact parameter vector
because of fluctuations that arise from having a finite number of nucleons.
For high-\pt\ particles, an azimuthal anisotropy can be induced if there
is stronger suppression of the hadron yield along the long axis than
the short axis of the overlap region.
The importance of jet-quenching measurements taking into account the
jet orientation relative to the geometry of the interaction region was 
first demonstrated
by the PHENIX experiment~\cite{Adare:2010sp}, where the azimuthal 
anisotropy of high-\pt\ neutral pions ($\pi^{0}$) was
studied up to $\pt \approx 18\GeVc$ in AuAu collisions
at \rootsNN\ = 200\GeV.

This Letter presents a study of the azimuthal anisotropy extended to
very high \pt\ (up to $\pt \approx 60\GeVc$) for PbPb
collisions at \rootsNN = 2.76\TeV at the LHC using the Compact Muon Solenoid (CMS) detector.
The $v_{2}$ coefficient is determined as a function of
charged particle \pt\ and overlap of the colliding nuclei (centrality)
in the pseudorapidity regions of $|\eta|<1$ and $1<|\eta|<2$,
where $\eta = -\ln [ \tan(\theta/2)]$ and
$\theta$ is the polar angle relative to the counterclockwise beam direction.
In the low momentum region (below a few \GeVc), $v_{2}$ is generally associated with
hydrodynamic flow~\cite{Voloshin:2008dg}, as distinct from the
jet energy-loss mechanism believed to dominate at high \pt (e.g., above 10\GeVc).
Using a single-track high-level trigger, a significantly larger event sample of high-\pt\ tracks
than previously available is obtained, providing the first precise
measurement of the $v_{2}$ coefficient above 20\GeVc
in heavy-ion collisions. Our results may impose quantitative constraints
on models of the in-medium energy loss of high-\pt\ partons, particularly
the influence of the path length and the shape of the interaction region
on the energy loss.

The data used in this analysis correspond to an integrated
luminosity of 150~$\mu$b$^{-1}$ and were recorded during the
2011 LHC heavy-ion
running period. A detailed description of the CMS
detector can be found in Ref.~\cite{CMS:2008zzk}. Its central feature is
a superconducting solenoid of 6~m internal diameter, providing a
magnetic field of 3.8~T. Within the field volume are the silicon
pixel and strip tracker, the crystal electromagnetic calorimeter,
and the brass/scintillator hadron calorimeter.
In PbPb collisions, trajectories of charged particles with $\pt > 200$\MeVc are
reconstructed in the silicon tracker covering the pseudorapidity
region $|\eta|<2.5$, with a track momentum resolution of about 1\% at \pt\ = 100\GeV/c.
In addition, CMS has extensive forward calorimetry, in
particular two steel/quartz-fiber \v{C}erenkov hadronic forward (HF)
calorimeters, which cover the pseudorapidity range $2.9 < |\eta| < 5.2$.
The HF calorimeters are segmented into towers, each of which is a two-dimensional
cell with a granularity of 0.5 unit in $\eta$ and 0.349 rad in $\phi$.

Minimum bias PbPb events were triggered by coincident
signals from both ends of the detector in either the beam
scintillator counters (BSC) at $3.23 < |\eta| < 4.65$ or in the HF calorimeters.
Events due to noise, cosmic rays, out-of-time triggers,
and beam backgrounds were suppressed by requiring a
coincidence of the minimum bias trigger with bunches colliding
in the interaction region. The trigger has an acceptance of $(97 \pm 3)$\%
for hadronic inelastic PbPb collisions.
Because of hardware limits on the data acquisition rate,
only a small fraction of all minimum bias triggered events were recorded.
To maximize the event sample with high-\pt\
particles emitted in the PbPb collisions, a dedicated
high-\pt\ single-track trigger was implemented using the CMS level-1
(L1) and high-level trigger system.
The trajectories of charged particles from the silicon tracker were found online
using a tracking algorithm identical to those employed in the offline track
reconstruction, starting with a candidate from the three-layer silicon pixel tracker
having $\pt>11$\GeVc. For PbPb events with total transverse energy at
L1 (L1\_ETT) above 100\GeV, a trigger efficiency
of above 95\% relative to offline reconstructed tracks
was achieved for track \pt\ greater than 12\GeVc. For the
events with L1\_ETT less than 100\GeV, an additional requirement of a calorimeter
jet at L1 with $\pt>16$\GeVc~\cite{Chatrchyan:2011sx} was
imposed in order to reduce the detector readout rate.
This resulted in an efficiency of about 75\% starting
at $\pt \approx 12\GeVc$ but increasing to almost 100\% above
$\pt \approx 20\GeVc$. In this analysis, the minimum bias data sample
is used for the $v_{2}$ measurement of tracks with $1<\pt<12$\GeVc
(also $12<\pt<20$\GeVc for cross-check purpose),
while the high-\pt\ track triggered sample is used in the range of $\pt>12$\GeVc.

Events are further selected offline by requiring energy deposits
in at least three towers in each of the HF calorimeters, with
at least 3\GeV of energy in each tower, and the presence of a reconstructed
primary vertex containing at least two tracks.
These criteria further reduce the
background from single-beam interactions (e.g., beam-gas and beam-halo),
cosmic muons, and large impact parameter, ultra peripheral
collisions that lead to the electromagnetic breakup of one or
both Pb nuclei~\cite{Djuvsland:2010qs}. The reconstructed
primary vertex is required to be located within $\pm$15~cm of the average
vertex position along the beam axis and within a radius of 0.02~cm
in the transverse plane. The centrality of the PbPb reaction is determined by taking
fractions of the total hadronic inelastic cross section, according to
percentiles of the distribution of the total energy deposited
in the HF calorimeters~\cite{Chatrchyan:2011sx}.
The centrality classes used in this analysis are 0--10\% (most central),
10--20\%, 20--30\%, 30--40\%, 40--50\%, and 50--60\%, ordered from the
highest to the lowest HF energy deposited.

The reconstruction of the primary event vertex and the
trajectories of charged particles in PbPb collisions
is based on signals in the silicon pixel and strip detectors and described
in detail in Ref.~\cite{cms:2012nt}. In selecting the charged primary tracks,
a set of tight quality selections were imposed to minimize the contamination
from misidentified tracks. These include requirements of a
relative momentum uncertainty of less than 5\%, at least 13 hits
on the track, a normalized $\chi^2$ for the track fit of less than 0.15 times the
number of hits, and transverse and longitudinal track displacements
from the primary vertex position less than three times the 
sum in quadrature of the measurement uncertainties. From
studies based on PbPb events simulated using \textsc{hydjet}~\cite{Lokhtin:2005px}
(version 1.6), the combined geometrical
acceptance and reconstruction efficiency of the primary tracks reaches about 66\% (50\%)
at $|\eta| < 1.0$ ($1.0 < |\eta| < 2.0$)
for the 5\% most central PbPb events, with little
dependence on \pt. For the peripheral PbPb events,
the efficiency is improved by up to 5\%, again
largely independent of \pt. The fraction of
misidentified tracks is kept at the 1-2\% level at $\pt>2\GeVc$
over almost the entire $\eta$ and centrality ranges.
It increases up to 10\% for very low \pt\ (${\approx}1\GeVc$) particles in
the forward ($|\eta| \approx 2)$ region for the 5\% most central PbPb events.

The analysis follows closely the event-plane method described in Ref.~\cite{Collaboration:2012ta}.
The observed $v_{2}$ value for a given centrality and \pt\
range is defined by $v_{2}^{obs}=\left\langle{\cos 2\left( {\phi - \Psi_2}\right)}\right\rangle$, where
the average is taken over all particles in all events within a centrality and \pt\
bin. The second-order ``event-plane'' angle $\Psi_2$ corresponds to the
event-by-event azimuthal angle of maximum particle
density in PbPb collisions. It is an approximation of the participant-plane angle $\Psi_\mathrm{PP}$,
which is not directly observable.
The determination of $\Psi_2$ uses the energy deposited in the HF calorimeters with
$\Psi_{2}  = \frac{1}{2}\tan ^{ - 1} \frac{{\sum\limits_i {w_i \sin \left( {2\phi _i } \right)} }}{{\sum\limits_i {w_i \cos \left( {2\phi _i } \right)} }}$, where the weight factor $w_i$ for
the $i^{th}$ tower at azimuthal angle $\phi_i$ is taken
as the corresponding transverse energy.
The sums are taken over all the towers within a slightly truncated
$\eta$ range of each HF calorimeter coverage. For the $v_{2}$ study in this analysis,
charged particles detected in the tracker with $\eta>0$ ($<0$) are correlated
with an event plane found using energy deposited in a region of
the HF spanning $-5 < \eta < -3$ ($3 < \eta < 5$). In this manner, a minimum $\eta$
gap of 3 units is guaranteed between particles used in the event-plane
determination and those for which the $v_2$ value is being measured,
thereby significantly reducing the effect of other correlations
that might exist, such as that from dijets.
This $\eta$ gap is particularly important for the high-\pt\ particle studies.

The resolution of the event plane depends on the centrality and is limited by
the finite number of particles used in its determination. The final
$v_2$ coefficient in the event-plane method is evaluated by dividing
the observed value $v_2^{obs}$ by a ``resolution-correction'' factor,
$R$, with $v_2=v_2^{\text{obs}}/R$ and where $R$ can range
from 0 to 1~\cite{Poskanzer:1998yz}, with
a better resolution corresponding to a larger value of $R$. 
An event-averaged resolution correction
factor can be found experimentally by extracting separate event-plane
angles using particles emitted into three non-overlapping $\eta$ regions.
In this  ``three-sub-event'' technique, which is described
in more detail in Ref.~\cite{Poskanzer:1998yz},
the resolution correction factor for a given $\eta$ region (denoted A,
with B and C used for the other two $\eta$ ranges) is found
using
$R_A=\sqrt{\frac{\langle \cos(2 (\Psi^A_{2}-\Psi^B_{2}))\rangle
\langle\cos(2 (\Psi^A_{2}-\Psi^C_{2}))\rangle}{\langle\cos(2
(\Psi^B_{2}-\Psi^C_{2}))\rangle}}$.
The averages are over all events corresponding to a given centrality
bin. Reconstructed tracks with $|\eta|<0.8$ and
$\pt>1$\GeVc are used for the ``third'' $\eta$ range (denoted C) introduced
to determine the resolution of the event planes found using the HF
detectors at $-5 < \eta < -3$ (denoted A) and $3 < \eta < 5$ (denoted B).
In the calculation of $\Psi^C_{2}$, the weight factor $w_i$ of the $i^\text{th}$ track at
angle $\phi_i$ is taken as the corresponding transverse momentum.
The extracted $R$ values for the HF event planes
are found to vary between 0.55 and 0.84, reaching a maximum
value for events in the 20--30\% centrality bin. The difference in the
resolution correction factors found for the two HF event planes is less than
$1\%$.

The sensitivity of the deduced $v_2$ values to the size of the
minimum $\eta$-gap is explored
by defining additional event planes with different pseudorapidity ranges.
Varying the gap size from 3 to 4 units,
the $v_{2}$ values are found
to be consistent within $\pm$2.5\% (central) or $\pm$10\% (peripheral).
For all systematic studies, relative uncertainties to $v_{2}$ are quoted.
The influence of misidentified tracks is another
source of systematic uncertainty at high \pt. The $v_{2}$ of misidentified
tracks in data is estimated using a very loose track selection. Taking the
misidentified-track $v_2$ signal, together with the
probability of reconstructing a fake track as determined from
simulated events, the systematic uncertainty on the
observed $v_{2}$ values is estimated to be $\pm$1--3\%, depending on \pt\
and centrality. Potential biases of the events triggered
by the high-\pt\ track algorithm are investigated. A comparison
of $v_{2}$ results for $12<\pt<20$\GeVc obtained from minimum bias
and high-\pt\ track triggered samples shows a variation of
less than $\pm$1\%. To account for non-uniformities in the detector acceptance that can lead to
artificial asymmetries in the event-plane angle
distribution and thereby also affect the deduced $v_2$ values, a 
Fourier-analysis-based ``flattening'' procedure is 
used~\cite{Poskanzer:1998yz} where each calculated event-plane angle 
is shifted slightly to recover
a uniform azimuthal distribution of angles~\cite{Collaboration:2012ta}.
Monte Carlo studies have shown that this flattening procedure fully
corrects for non-uniformities in the CMS detector response.
The overall systematic uncertainties associated with the event-plane flattening
procedure and the resolution correction determination are found to be less
than $\pm$1\%. Additional uncertainties due to track-quality requirements are
examined by loosening or tightening the selection criteria described previously
and are found to be less than $\pm$ 0.5\%. The different
systematic sources are added in quadrature
to obtain the overall systematic uncertainty in $v_{2}$.

\begin{figure}[th!]
  \begin{center}
    \includegraphics[width=\linewidth]{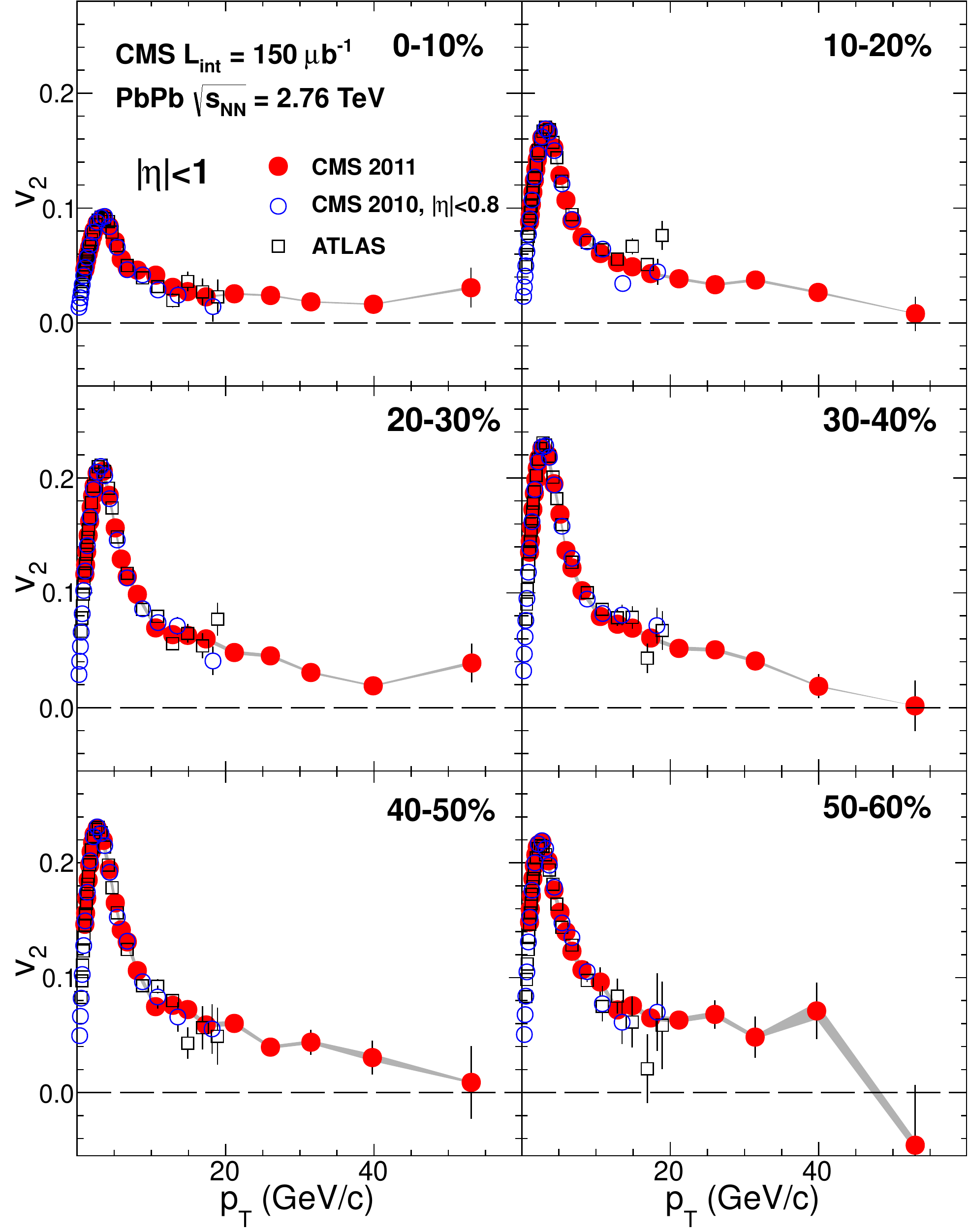}
    \vspace{-5mm}
    \caption{The single-particle azimuthal anisotropy, $v_{2}$,
    as a function of the charged particle transverse momentum
    from 1 to 60\GeVc with $|\eta|<1$ 
    for six centrality ranges in PbPb collisions
    at \rootsNN\ = 2.76\TeV, measured by the CMS experiment (solid markers).
    Error bars denote the statistical uncertainties, while the grey bands
    correspond to the small systematic uncertainties.
    Comparison to results from the ATLAS (open squares) and
    CMS (open circles) experiments using data collected in 2010 is also shown.}
    \vspace{-2mm}
    \label{fig:v2_pt_ep_atlas_mid}
  \end{center}
\end{figure}

The results of $v_{2}$,
as a function of \pt\ from 1 to 60\GeVc for events with centralities
ranging from 0--10\% (central) to 50--60\% (peripheral),
are presented in Fig.~\ref{fig:v2_pt_ep_atlas_mid}
for $|\eta|<1$. 
The highest \pt\ bin covers a range from 48.0 to 60.8\GeV/c. The new CMS results in this paper
significantly extend the \pt\ reach of $v_{2}$ measurements previously achieved by
the ALICE~\cite{Aamodt:2010pa}, ATLAS~\cite{ATLAS:2011yk} and CMS~\cite{Collaboration:2012ta}
collaborations.
The CMS and ATLAS results based on event-plane analyses
using forward calorimeters for the event-plane determination
show agreement within 2--3\% up to $\pt \approx 20\GeVc$.

The $\pt$-dependence of $v_{2}$ shows a 
trend of rapid rise, reaching a maximum at $\pt \approx 3\GeVc$,
followed by a dramatic decrease for \pt\ values up to about 10\GeVc.
Beyond $\pt \approx 10\GeVc$, the $v_2$ values show a much 
weaker dependence on \pt and gradually decrease, but
remain larger than zero up to at least \pt\ $\approx$ 40 \GeVc. 
The magnitude of $v_2$ at high \pt\ can probe the 
path length ($l$) dependence of parton energy loss ($\Delta E$), $\Delta E \sim l^{\alpha}$~\cite{Peigne:2008wu,Wicks:2005gt,Jia:2010ee,Jia:2011pi,Renk:2010mf,Betz:2011tu}. 
At RHIC energies, the $\pi^{0}$ $v_2$ data for
\pt\ $\sim$ 6--10\GeVc support a scenario of $\alpha=3$ based on
AdS/CFT gravity-gauge dual modeling, while the 
pertabutive QCD calculations ($\alpha=1$ for collisional 
energy loss and $\alpha=2$ for radiative energy loss) seem to 
underpredict the data~\cite{Adare:2010sp}. 
The $v_2$ results reported in this paper extends to a
significantly wider \pt\ regime, where particle production
is unambiguously dominated by parton fragmentation and energy-loss
models are expected to be more applicable.
The clear $\pt$-dependence of $v_2$ observed at the LHC energy suggests
that additional dynamical modeling of parton modification, particularly 
by incorporating the proper initial parton energy dependence, is required in 
order to further examine the path length dependence of energy loss 
(i.e., $\alpha$ parameter).

\begin{figure}[th!]
  \begin{center}
    \includegraphics[width=\linewidth]{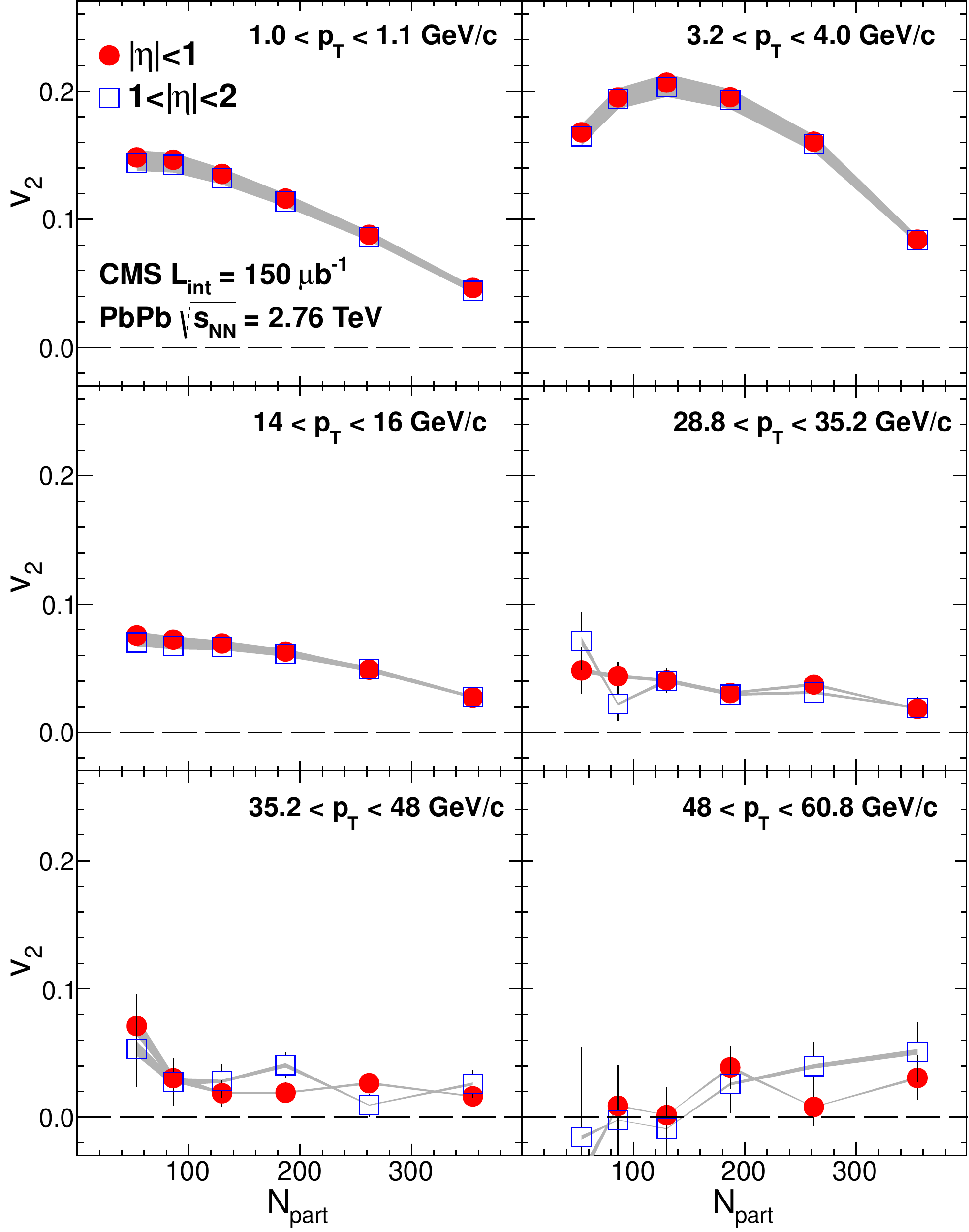}
    \vspace{-5mm}
    \caption{The single-particle azimuthal anisotropy, $v_{2}$,
    as a function of the number of participating nucleons with $|\eta|<1$ (solid circles)
    and $1<|\eta|<2$ (open squares) for six selected \pt\ ranges
in PbPb collisions
    at \rootsNN\ = 2.76\TeV, measured by the CMS experiment.
    Error bars denote the statistical uncertainties, while the grey bands
    correspond to the systematic uncertainties.}
    \label{fig:v2_npart_ep}
    \vspace{-4mm}
  \end{center}
\end{figure}

Figure~\ref{fig:v2_npart_ep} shows the $v_{2}$ dependences 
on the number of participant nucleons (\npart) associated with each
centrality bin through a Glauber model calculation. The corresponding
participant eccentricities of the the overlap region vary from 0.46 to 0.093.  
Different \pt\ bins are
shown for two pseudorapidity ranges, $|\eta|<1$ and $1<|\eta|<2$. The results 
appear to be independent of $\eta$ in all \pt\ bins
within the statistical uncertainties. Also, the $v_2$ values tend to decrease
with increasing collision centrality (i.e., larger \npart)
over a wide \pt\ range (although this trend appears to be reversed
for $3.2<\pt<4.0$\GeVc toward very peripheral events). 
This behavior is expected for low-\pt\ (below a few \GeVc) particles in the
context of the relationship between hydrodynamic
flow phenomena and the eccentricity of the initial-state collision geometry.
The similar trend observed for very high-\pt\ particles, at least up to \pt\ $\approx 48$\GeVc, 
reflects how the $v_2$ results at high \pt\ are also sensitive to the initial geometry.
This indicates that the initial conditions of the hot QCD medium can be 
further constrained by simultaneously comparing data with theoretical 
calculations from both hydrodynamics at low \pt\ and in-medium parton energy loss at high $\pt$.

In summary, the azimuthal anisotropy of charged particles with respect to
the event plane has been studied in PbPb collisions at \rootsNN\
= 2.76\TeV using the CMS detector. The $v_{2}$ coefficient
was determined over a wide range in \pt\ up to approximately
60\GeVc, as a function of collision centrality.
The results reported in this paper significantly improve the
statistical precision of previous $v_{2}$ measurements for $12<\pt<20$\GeVc,
and explore for the first time the very high \pt\ region beyond 20\GeVc.
The $v_{2}(\pt)$ behavior shows a trend of rapid rise to
a maximum at $\pt \approx3\GeVc$ and a subsequent fall for all centrality
and $|\eta|$ ranges. Beyond $\pt \approx 10\GeVc$, the observed $v_{2}$
values still show a clear \pt\ dependence but with a more moderate decrease
with \pt, remaining finite up to at least $\pt\approx 40\GeVc$.
A common trend in the centrality dependence of $v_{2}$ is observed
for particles over a wide range of \pt\ up to approximately 48\GeVc,
suggesting a potential connection to the initial-state geometry.
The precision data over a wide kinematic range presented here will
provide important constraints on models of parton energy loss, particularly
in terms of its dependence on the initial conditions, parton energy and path length
through the medium.

We wish to congratulate our colleagues in the CERN accelerator departments for the excellent performance of the LHC machine. We thank the technical and administrative staff at CERN and other CMS institutes, and acknowledge support from: FMSR (Austria); FNRS and FWO (Belgium); CNPq, CAPES, FAPERJ, and FAPESP (Brazil); MES (Bulgaria); CERN; CAS, MoST, and NSFC (China); COLCIENCIAS (Colombia); MSES (Croatia); RPF (Cyprus); Academy of Sciences and NICPB (Estonia); Academy of Finland, MEC, and HIP (Finland); CEA and CNRS/IN2P3 (France); BMBF, DFG, and HGF (Germany); GSRT (Greece); OTKA and NKTH (Hungary); DAE and DST (India); IPM (Iran); SFI (Ireland); INFN (Italy); NRF and WCU (Korea); LAS (Lithuania); CINVESTAV, CONACYT, SEP, and UASLP-FAI (Mexico); MSI (New Zealand); PAEC (Pakistan); SCSR (Poland); FCT (Portugal); JINR (Armenia, Belarus, Georgia, Ukraine, Uzbekistan); MON, RosAtom, RAS and RFBR (Russia); MSTD (Serbia); MICINN and CPAN (Spain); Swiss Funding Agencies (Switzerland); NSC (Taipei); TUBITAK and TAEK (Turkey); STFC (United Kingdom); DOE and NSF (USA).
\bibliography{auto_generated}   
\cleardoublepage \appendix\section{The CMS Collaboration \label{app:collab}}\begin{sloppypar}\hyphenpenalty=5000\widowpenalty=500\clubpenalty=5000\textbf{Yerevan Physics Institute,  Yerevan,  Armenia}\\*[0pt]
S.~Chatrchyan, V.~Khachatryan, A.M.~Sirunyan, A.~Tumasyan
\vskip\cmsinstskip
\textbf{Institut f\"{u}r Hochenergiephysik der OeAW,  Wien,  Austria}\\*[0pt]
W.~Adam, T.~Bergauer, M.~Dragicevic, J.~Er\"{o}, C.~Fabjan, M.~Friedl, R.~Fr\"{u}hwirth, V.M.~Ghete, J.~Hammer\cmsAuthorMark{1}, N.~H\"{o}rmann, J.~Hrubec, M.~Jeitler, W.~Kiesenhofer, M.~Krammer, D.~Liko, I.~Mikulec, M.~Pernicka$^{\textrm{\dag}}$, B.~Rahbaran, C.~Rohringer, H.~Rohringer, R.~Sch\"{o}fbeck, J.~Strauss, A.~Taurok, F.~Teischinger, P.~Wagner, W.~Waltenberger, G.~Walzel, E.~Widl, C.-E.~Wulz
% ...
\vskip\cmsinstskip
\dag:~Deceased\\
1:~~Also at CERN, European Organization for Nuclear Research, Geneva, Switzerland\\
% ...

\end{sloppypar}
\end{document}